\documentclass[12pt]{article}
\usepackage{amsmath}
\usepackage{amsfonts}
\usepackage{amssymb}
\usepackage{amsthm}
\usepackage{stackrel}

\usepackage{color}

\usepackage{figsize}

\usepackage{graphics}

\newtheorem{ex}{Example}

\newtheorem{thm}{Theorem}[section]

\newcommand{\R}{{\rm I}\kern-0.18em{\rm R}}
\newcommand{\1}{{\rm 1}\kern-0.25em{\rm I}}
\newcommand{\E}{{\rm I}\kern-0.18em{\rm E}}
\newcommand{\p}{{\rm I}\kern-0.18em{\rm P}}
\makeatletter

\title{One look at the rating of scientific publications and corresponding toy-model}
\author{Lev B. Klebanov\footnote{Department of Probability and Mathematical Statistics, Charles University, Czech Republic}}
\date{}
\begin{document}

\maketitle

\section{Introduction}\label{sec1}
\setcounter{equation}{0}

It is well-known fact that the distribution of scientific production has hyperbolic form. One of the descriptions for this distribution is given by celebrated Lotka's law \cite{Lot} (more details about this law see for example in \cite{Vla}). Corresponding distribution has heavy (power with exponent 1) tail. However, it is interesting to understand what is the reason for it to have heavy tail. Is it connected to different scientific abilities, or is it explained by random character of the citation process? Our aim in this paper is not to give full explanation of the heavy tail effect and describe the distribution of scientific publication number or of that of citations. We try only to explain that the second reason (randomness of corresponding process) cannot be ignored. That is why many models of scientific production distribution are not mentioned here.

In the latest time there exists a tendency to estimate scientific results of an scientist or ``prestige" of a journal basing on their impact factor which is based on the citations number of corresponding author(s) or journal(s). This approach obviously is based on a naive idea that the number of citations tells us something essential about scientific value of cited publication. However, it is not clear how essential is this information. For example, one cannot say is the difference between the number of citations for two different scientists just random or it says that their impact in scientific field is essentially different. To understand this one needs to have mathematical model of publication and citation process. Our aim in this paper is to propose corresponding toy-model, that is a model which takes into account some properties of the publication-citation process, and see what type of consequences provide these properties.

\section{Toy-model for scientific citation model and the rating of scientific publications}\label{sec2}
\setcounter{equation}{0}

We shall start with very simple model of scientific papers citation. The model represented in this section is a slight modification of that proposed in \cite{KS}. 

Suppose that an author has at least one publication (in other case there is nothing to cite). And suppose the author sends manuscripts for further publication to different journals. Let us make some assumptions, which will be discussed later.

{\bf Assumption 1.} {\it Let the probability of rejection of each manuscript be $q$, and the decisions on each manuscript publication are taken independently}.

Then the probability for this scientist to have exactly $k$ publications equals to  $q(1-q)^{k}$. This means, the number of published papers has geometric distribution. Its probability generating function (p.g.f.) is
\[Q(z) = \frac{q}{1-(1-q)z}.\]
Of course, here we suppose that all journals the author is sending manuscripts for publication have the same referee system, that is they accept submissions with the same probability $1-q$. More realistic is the situation with different (random) parameter $q$:
\[  Q(z) = \int_{0}^{1}\frac{q}{1-(1-q)z}d\,\Xi(q).\] 

It is naturally to suppose that each publication generates a number (may be zero) of citations. Of course, the probability for a paper to be cited again depends on the amount of previous citations. 

{\bf Assumption 2.} {\it Denote by $a$ the probability that a paper will have no citations at all. Suppose too, that the probability that the paper having $k-1$ ($k \geq 1$) citation will have no other citations is $p/k$, where $p$ is the probability that the paper will not be cited again after the first citation.}

Therefore the probability that a paper will be cited exactly $k$ times is 
\[ a+(1-a)\frac{p}{k}\prod_{j=1}^{k-1}(1-p/j).\] 
Therefore, the p.g.f. of the distribution for citations number of this paper is
\[ a+(1-a)(1-(1-z)^p)= 1- (1-a)(1-z)^p. \]
The p.g.f. of the citations number for all papers coming from one author is the superposition of the latest p.g.f. and that of geometric distribution, that is
\begin{equation}\label{eq1}
{\mathcal P}(z)=1-(1-a)\Bigl(1-\frac{q}{1-(1-q)z}\Bigr)^p.
\end{equation}
This distribution has parameters $a \in [0,1]$, $q \in (0,1]$, and $p \in (0,1]$. For the case $a<1$, $p<1$ this distribution has heavy tail. The mean value is infinite in this case.

If we are interested in the distribution of the number of citations in a field of science we suppose the following. 

{\bf Assumption 3} {\it We assume that the number of all scientists publishing paper in the field has Poisson distribution with parameter $\lambda$.} 

The number of all citations in this field of science has p.g.f.
\begin{equation}\label{eq2}
{\mathcal R}(z) = \exp\Bigl\{-\lambda (1-a)(1-q)^p\Bigl(\frac{1-z}{1-(1-q)z}\Bigr)^p\Bigr\} .
\end{equation}
Note that this p.g.f. belongs to the class of discrete stable distribution with ``normalizing" p.g.f. 
\begin{equation}\label{eq3}
{\mathcal Q}(z) = \frac{(1-u)+(u+q-1)z}{1-u(1-q)-(1-q)(1-u)z}, 
\end{equation}
where $u \in (0,1)$ is a parameter.

Obviously, the distribution with p.g.f. (\ref{eq2}) has heavy tail equivalent (up to constant multiplier) to that of (\ref{eq1}). As it was noted above, for the case of $a<1$ and $q<1$ the mean is infinite. This distribution has the mode at zero and finite positive median. Therefore, for any finite set of observations on (\ref{eq2}) (or on (\ref{eq1})) the empirical mean is much larger that the empirical median. Also, large number of citations will correspond to a relatively small amount of publications, while the main part of the papers will have a small number of citations. This effect is not connected to any asymptotic result. In our toy-model no essential difference between scientists (or the quality of their papers) is supposed. The noticeable difference between the numbers of citations is attributed to the random nature of publications and citation processes. 
The model proposed above is in contradiction with the main idea of ranking scientists, journals or scientific organizations on the base of the number of citations or on corresponding impact factor. According to the model, including the scientists having many citations into scientific elite group is equivalent to include the people who won the lottery into the group of successful businessmen. 

\section{Analysis of the assumptions of toy-model}\label{sec3}
\setcounter{equation}{0}

Let us start with {\bf Assumption 1}. If scientist has a fixed quality of his/her manuscripts, all the results are new and interesting on the same level, then the probability of (random) rejection of each manuscript seems to be the same for all manuscripts. Of course, with time the quality of the papers changes. However, for a fixed (not too long) time interval this assumption seems to be a good approximation to reality for the case that the number of attempts to publish (that is, the number of papers submitted) is bounded. This leads to truncated geometric distribution with p.g.f.
\[ Q_m(z) = \frac{q(1-(1-q)^m z^m)}{(1-(1-q)^m)(1-(1-q)z)},\] 
where $m$ is the number of attempts. It is clear that the case of large $m$ is very similar to that when $m \to \infty$. Therefore, Assumption 1 seems to be rather natural and not very restrictive. Of course, the variant of Assumption 1 with random parameter $q$ is more realistic, but difficult for analysis. 

Of course, Assumption 1 does not take into account the time passing between applications for publications. Therefore, it ignores any changes in the author status which may affect referee opinion and/or stile of writing and choice of papers thematic. All this means that Assumption 1 contains essential simplifications of reality.

{\bf Assumption 2} looks less natural than the first one. Firstly, we have to explain why the probability of $k$th citation of the paper is higher that for the case of previous citation. 
This follows from the results by Simkin \& Roychowdhury (\cite{SR5}). They wrote: ``Recent scientific research points to the evidence that the majority of scientific citations were not read by the citing authors. Apart from the analysis of misprint propagation this conclusion is indirectly supported by recent study, which
found that the correlation coefficient between the number of citations to and the number of readings of papers in arXiv.org is only $r \sim 0.45$. This suggests that just 20\% ($r^2 \sim 0.2$) of variance in number of citations is explained by the variance in the number of readings." Let us also note the paper \cite{SR3} containing similar conclusions. 

So, we see that a lot of citations are just copied from previous publications and therefore, the probability of such copying is higher for the papers with larger amount of citations. 

We cannot prove that the probability of $k$th citation will be proportional to $p/k$. However, this is a very simple (one of the most simpleast) dependence on $k$. Nothing show we cannot try this type of dependence and look at corresponding consequences. Similar to the case of Assumption 1, it would be better to have a bounded number of citations, that is come to truncated variant of the distribution (\ref{eq1}). It will lead to the change of tail behavior for corresponding distribution. However, its main body will be essentially the same, and one will observe almost the same behavior of empirical data.

{\bf Assumption 3} seems to be not very restrictive in view of known properties of Poisson distribution.

\section{Another approach to citation process  modeling}\label{sec4} 
\setcounter{equation}{0}

Here we represent another way to find approximation for citations number distributions under mild priory assumptions. Unlike to Section \ref{sec2}, the results here have asymptotic character and therefore are approximative. 

Our first assumption here is the same as Assumption 1 of Section \ref{sec2}. Namely, the number of published papers is supposed to have geometric distribution with probability generating function
\begin{equation}\label{eq4}
Q(z) = \frac{q}{1-(1-q)z},\qquad q\in (0,1). 	
\end{equation}
Each published paper either has no citations at all with the probability $r \in (0,1)$ or, with probability $1-r$, has a random number of citations. Suppose that the distribution of this citations number has p.g.f. $\mathcal{S}(z)$. If there is a group of $n$ such authors working independently and having the same distribution of their scientific production then the citations number of this group has p.g.f. of the form
\begin{equation}\label{eq5}
\mathcal{R}_n(z;r)=\mathcal{S}^n(r+(1-r)Q(z)).
\end{equation}
To understand limit behavior of (\ref{eq5}) we may apply a limit theorem to $\mathcal{S}^n(r+(1-r)z)$ as $r \to 1$ and $n \to \infty$.

\begin{thm}\label{th1} Suppose that\footnote{Symbol $\sim$ below denotes that corresponding ratio tends to $1$.} 
\begin{equation}\label{eq6} 
1-\mathcal{S}(z) \sim \lambda (1-z)^{\gamma} \;\; \text{as} \;\; z\to 1 
\end{equation}
for some $\lambda>0$ and $0<\gamma \leq 1$. Then
\begin{equation}\label{eq7} 
\mathcal{R}_{n}(z,r_n) \to \exp\Bigl\{-\lambda (1-q)^{\gamma}\Bigl(\frac{1-z}{1-(1-q)z}\Bigr)^{\gamma}\Bigr\}, \;\; \text{as} \;\; n \to \infty, 
\end{equation}
where $r_n \sim n^{-1/\gamma}$.
\end{thm}
Let us mention that this Theorem is known. See, for example \cite{CS}. However, we give a very simple proof here.
\begin{proof} 
From (\ref{eq6}) it follows that 
\[ \mathcal{S}^n(r_n+(1-r_n)z) \sim \bigl(1-\lambda (1-r_n)^{\gamma})(1-z)^{\gamma}\bigr)^n \longrightarrow \exp \{-\lambda (1-z)^{\gamma}\} \]
as $n \to \infty$. To finish the proof it is sufficient to substitute (\ref{eq4}) into previous relation.
\end{proof}
Let us mention again that the p.g.f.
\begin{equation}\label{eq7}
\mathcal{P}(z)=\exp\Bigl\{-\lambda (1-q)^{\gamma}\Bigl(\frac{1-z}{1-(1-q)z}\Bigr)^{\gamma}\Bigr\}
\end{equation}
belongs to the class of discrete stable distribution having ``normalizing" function (\ref{eq3}) (see \cite{KS}). Properties of corresponding distribution were studied in details by L. Sl\'{a}mov\'{a} \cite{Sl}. Here we mention that this distribution has heavy (power with index less than $1$) tails for the case $\gamma \in (0,1)$. In the case $\gamma = 1$, $q \in (0,1)$ the tails are exponential of index $q$, so that for smaller values of $q$ the tails are more heavy. For the case $\gamma =1$ and $q=1$ we have Poisson distribution.  

\section{One simple approach to scientific production process based on difference between scientists}\label{sec5} 
\setcounter{equation}{0}

Our first assumption here is similar to Assumption 1 of Section \ref{sec2}. Namely, the number of published by one author papers is supposed to have geometric distribution with probability generating function (\ref{eq4}). However, we suppose that the scientists have different abilities and, therefore, each scientist has her/his own parameter $q$ in (\ref{eq4}). For a group of such scientists the parameter $q$ may be considered as random with some (unknown) distribution $\Xi (q)$ in $(0,1)$ interval. Then p.g.f. of the number of papers published by this group of scientists is
\begin{equation}\label{eq8} 
Q(z)=\int_{0}^{1}\frac{q}{1-(1-q)z}d\,\Xi(q).
\end{equation}
Suppose that for some $A>0$ and $s \in \R^1$
\begin{equation}\label{eq9}
\Xi(\varepsilon) \sim A \varepsilon^s, \;\; \text{as}\;\; \varepsilon \to 0.
\end{equation}
Differentiating formally (\ref{eq8}) under integral sign $k$ times and passing to limit as $z \to 1$ we obtain
\begin{equation}\label{eq10}
Q^{(k)}(1) = k! \int_{0}^{1}\frac{(1-q)^k}{q^k}d\,\Xi (q).
\end{equation}
From (\ref{eq10}) it is easy to see that $k$th moment of $Q$ does not exists for $k>s$. It shows that the tails of $Q$ are rather heavy.
If the process of citation has the same probability structure for different journals we may finish model construction similarly to previous sections. In this case 
\begin{equation}\label{eq11} \mathcal{P}(z)=\exp\Bigl\{-\lambda \Bigl(\int_{0}^{1}\frac{(1-q)(1-z)}{1-(1-q)z}d\,\Xi (q)\Bigr)^{\gamma}\Bigr\}. 
\end{equation}
Under condition (\ref{eq9}) the distribution with p.g.f. (\ref{eq11}) has more heavy tails than for the case of (\ref{eq7}). However, the asymptotic  (\ref{eq9}) uses the fact that there are scientists having arbitrary small values of the parameter $q$, that is having arbitrary intensive productivity. This seems to be non-realistic. In the case when support of $\Xi$ is bounded out of zero the tail character is similar to that of the distribution for previous model. 

\section{Is it possible to distinguish models from sections \ref{sec4} and \ref{sec5}? }\label{sec6}
\setcounter{equation}{0}

For simplicity, let us call the model from Section \ref{sec4} as ``Equality model" because all scientist have the same distribution in the sense of their scientific production. The model from Section \ref{sec5} is called ``Elite model" in view of the fact that there are very productive scientists (elite). We are interested to answer the following questions: 
\begin{enumerate} 
\item[Q1.] Is it possible basing on empirical data distinguish ``Equality" and ``Elite" models? 
\item[Q2.] Is it possible to estimate scientific value of a paper or of an author basing on citation information among chosen model?
\item[Q3.] What is a connection of such indexes as Impact Factor of journal, Hirsh index or i:10 index with above mentioned models and/or scientific value of a paper?
\end{enumerate}

Let us start with Q1. The answer is affirmative for the case of non-random parameter $q$ in ``Equality" model. Really, for this case, the difference between ``Equality" and ``Elite" models lies in the non-degenerate character of $\Xi$ distribution for ``Elite" model\footnote{Let us mention that ``Equality" model is a particular case of ``Elite" for degenerate distribution $\Xi$}. Therefore, it is sufficient to test whether the variance of a characteristic of random parameter $p$ is strictly positive. 

However, mathematical expressions for both models are identical for the case of random $q$-parameter character in ``Equality" model. Therefore, one cannot distinguish between models for this general case. 

Let us pass now to the question Q2. Citation process for both models above is supposed to be random and independent on scientific value of any paper. Therefore, the answer to Q2 is negative. Of course, one may hope there are some other models for which the answer will be affirmative.
However, to proceed in this direction one needs not only have such model but also prove, the ours models do not correspond to reality. As far as it is known to the authors of this paper, there are neither such mathematical models no corresponding proofs.

Turn now to the question Q3. Our both models do not take into account any difference between journal quality, their circulation, or other circumstances closed to the mentioned. Therefore, the models proposed above have no connection to Impact Factor of a journal. We did not like to include in model construction any elements reflecting Impact Factor. One of the reasons for that is the main conclusion of \cite{LLG}. Namely, the authors of \cite{LLG} wrote: `` Since 1990, the proportion of highly cited papers coming from highly cited journals has been decreasing, and accordingly, the proportion of highly cited papers not coming from highly cited journals has also been increasing. Should this pattern continue, it might bring an end to the use of the Impact Factor as a way to evaluate the quality of journals, papers and researchers."

\section{Real data analysis. Publications number}\label{sec7}
\setcounter{equation}{0}

Here we consider some data on the number of publications made by members of some mathematical departments of universities. The data are taken from Internet site www.researchgate.net on 20.05.2017. We ignore such databases like Web Of Sciences and Scopus because they  give information about peer-reviewed literature only and therefore, are highly incomplete. Reviewing procedures will be considered a little bit later. Of course, we cannot guarantee completeness of information provided by Research Gate. This means our considerations do not provide correctness of our models. However, we proposed toy-models only and we do not need any proof for them, we are looking for some supporting data only.

\begin{ex}\label{ex1} 
Let 
\begin{align*}
T = \{130, 4, 27, 9, 12, 36, 32, 19, 129, 1,\\ 167, 278, 41, 46, 26, 25, 19, 12, 7, 11, 6, 2\}
\end{align*}
represent non-zero numbers of publications by members of Department of Probability and Mathematical Statistics Charles University \footnote{We take persons mentioned as such members on the site. However, not all real members of the Department were included, and, opposite, some non-members were included.}. 
\begin{figure}[htp]
	\centering
	\hfil
	\includegraphics[scale=0.9]{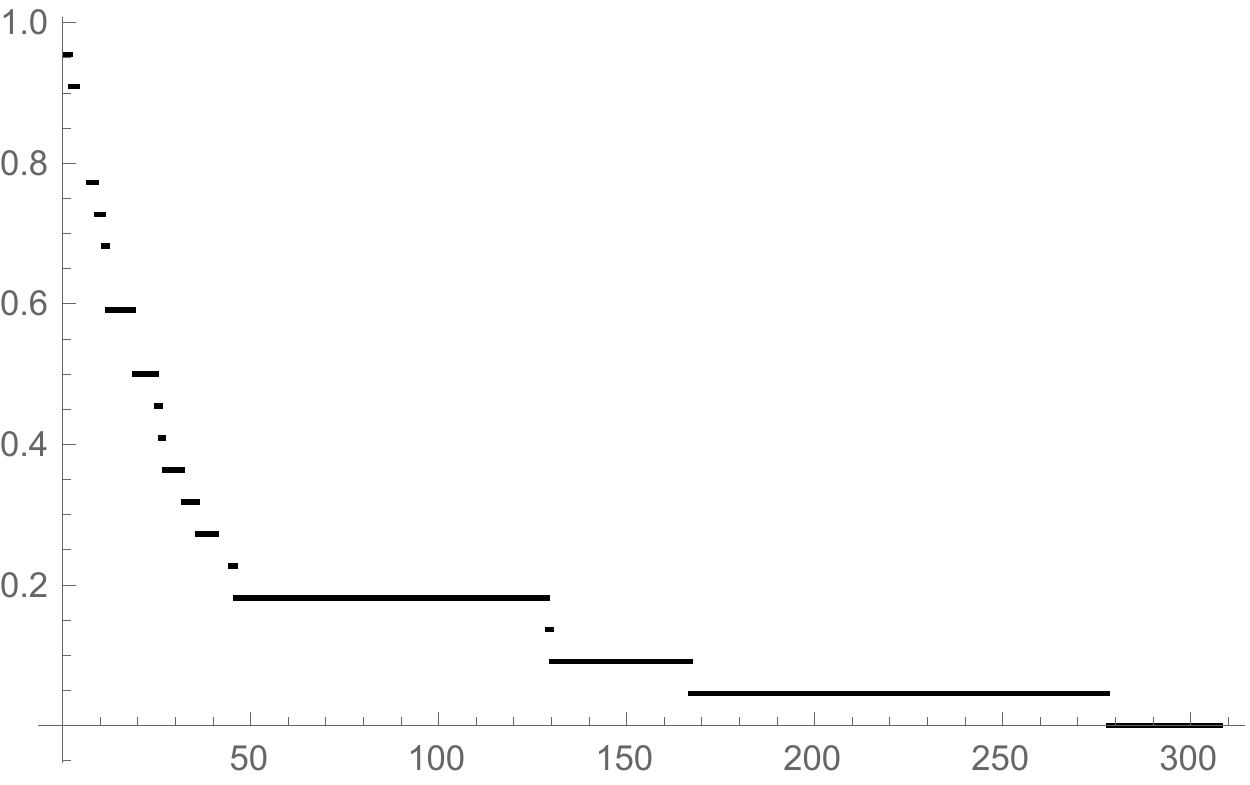}
	\caption{Plot of $1-F(x)$, where $F(x)$ is empirical distribution of the number of publications for Example \ref{ex1}.}\label{fig1}
\end{figure}

\begin{figure}[htp]
	\centering
	\hfil
	\includegraphics[scale=0.9]{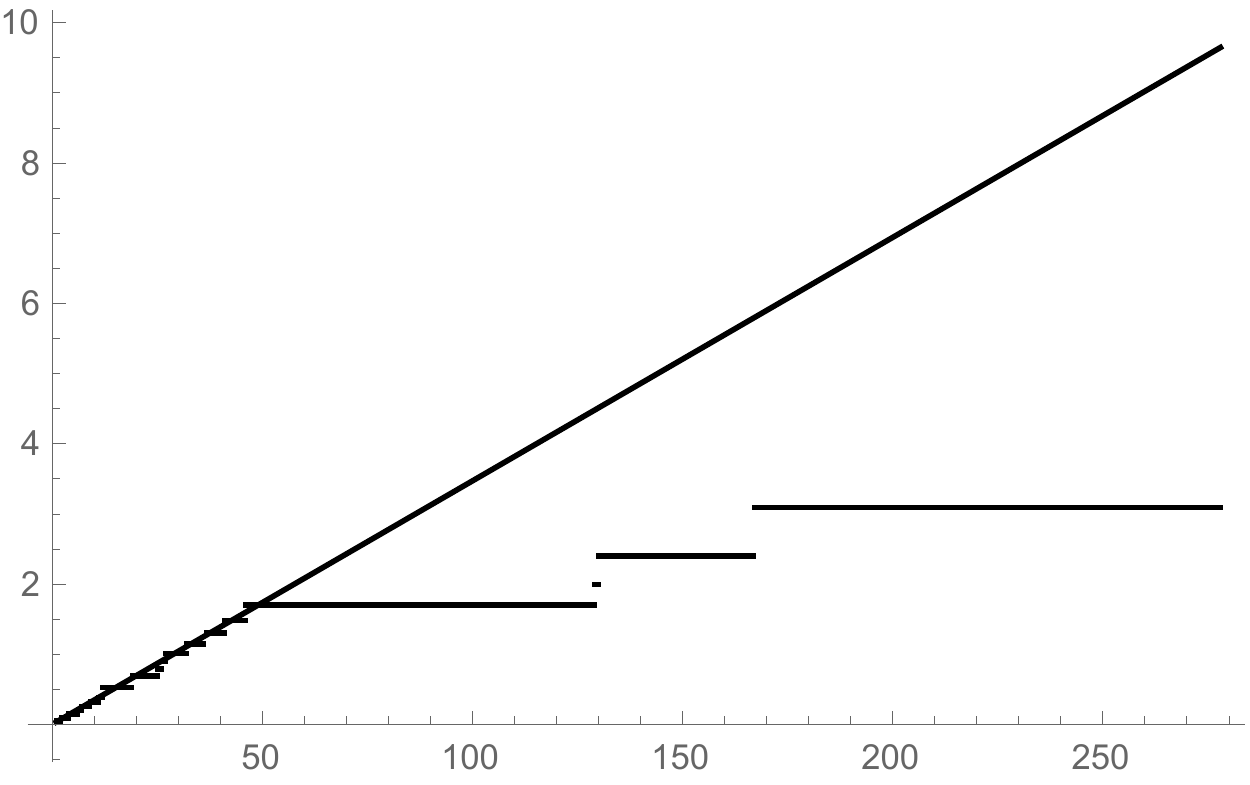}
	\caption{Plot of $-\log(1-F(x))$, where $F(x)$ is empirical distribution of the number of publications for Example \ref{ex1}. Straight line is an approximation for small number of publications (not greater that $50$).}\label{fig2}
\end{figure}

\begin{figure}[htp]
	\centering
	\hfil
	\includegraphics[scale=0.9]{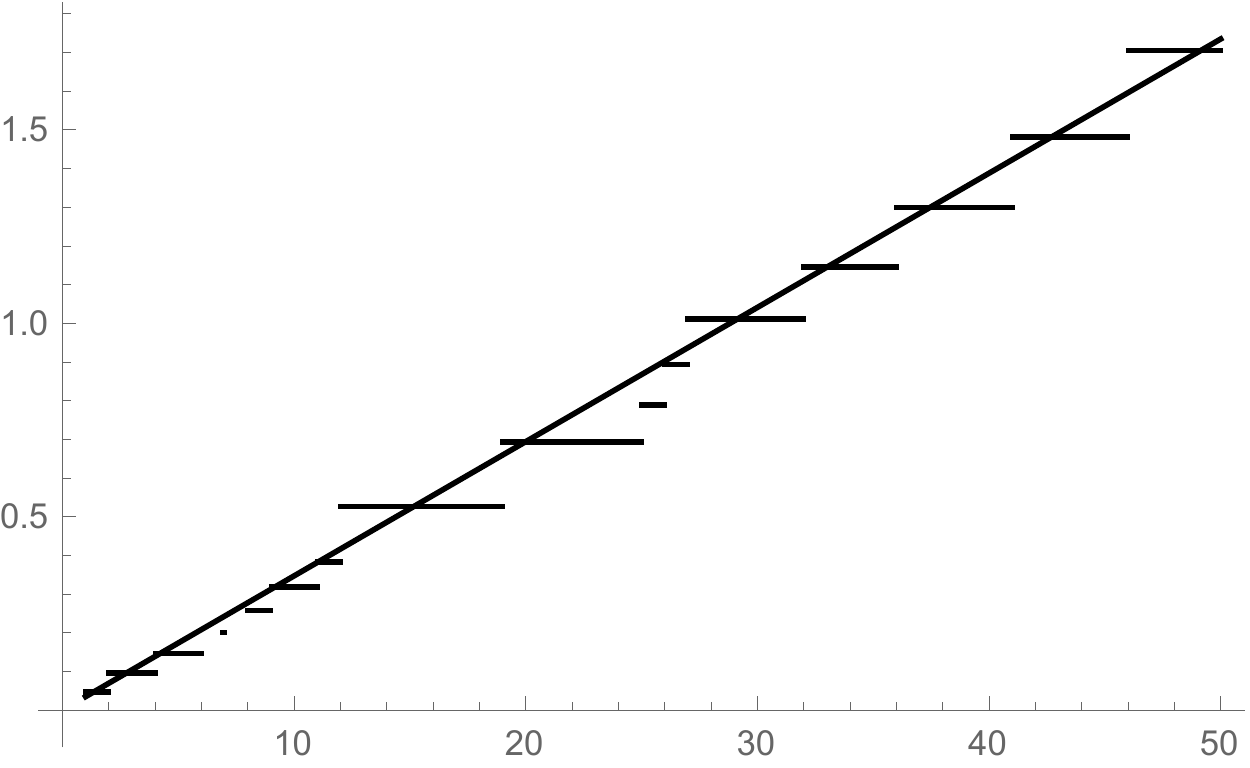}
	\caption{Plot of $-\log(1-F(x))$ for small number of publications (not greater that $50$) separately.}\label{fig3}
\end{figure}

If the number of publications has geometric distribution (this is Assumption 1 from Section \ref{sec2}) then $-\log(1-F(x))$ must have good approximation by a straight line. On Figures \ref{fig2} and \ref{fig3} we see not bad agreement between $-\log(1-F(x))$ and straight line for not large values of $x$. This fact supports Assumption 1 at least for not very large numbers of publications. On Figure \ref{fig2} we see that there are persons having abnormally large numbers of publication. This fact may be explained in many ways. Some of the reasons for that may be, for example, such: 
\begin{enumerate}
\item High qualification and experience in this field of science.
\item Random character of the parameter $q$ from Assumption 1 (in connection, say, with different type of journals the papers were published in).
\item Different conditions and interest to this field of science during rather long time interval passing between first and last publications.
\end{enumerate}
\end{ex} 

Example \ref{ex1} shows us that toy-model from Section \ref{sec2} may not be completely rejected and the difference between numbers of  publication does not show any difference of their scientific abilities. However, we may doubt this is a general situation. To be more sure we need more examples of similar situation. 

\begin{ex}\label{ex2} Let $	T = \{25, 76, 173, 2, 10, 9, 4, 13, 23\}$ represent non-zero numbers of publications by members of Department of Probability and Mathematical Statistics Saint-Petersburg State University (data taken from Research Gate 20.05.2017). On Figures \ref{fig4} and \ref{fig5} we see plots of $-\log (1-F(x))$ for all observations and for $x< 30$ correspondingly. $F(x)$ is corresponding empirical distribution function for the number of publications. 

\begin{figure}[htp]
	\centering
	\hfil
	\includegraphics[scale=0.65]{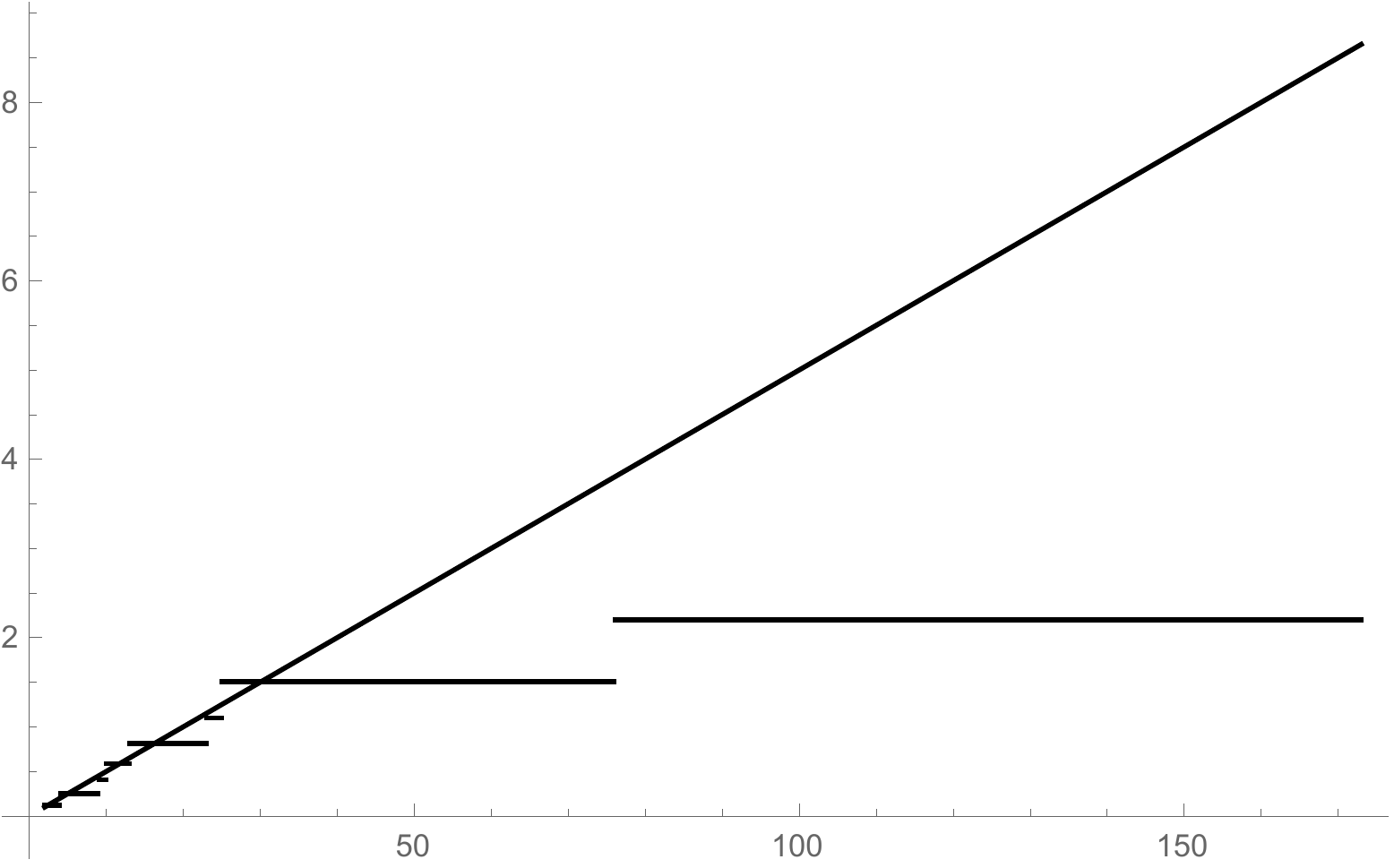}
	\caption{Plot of $-\log(1-F(x))$.}\label{fig4}
\end{figure}
	
\begin{figure}[htp]
	\centering
	\hfil
	\includegraphics[scale=0.65]{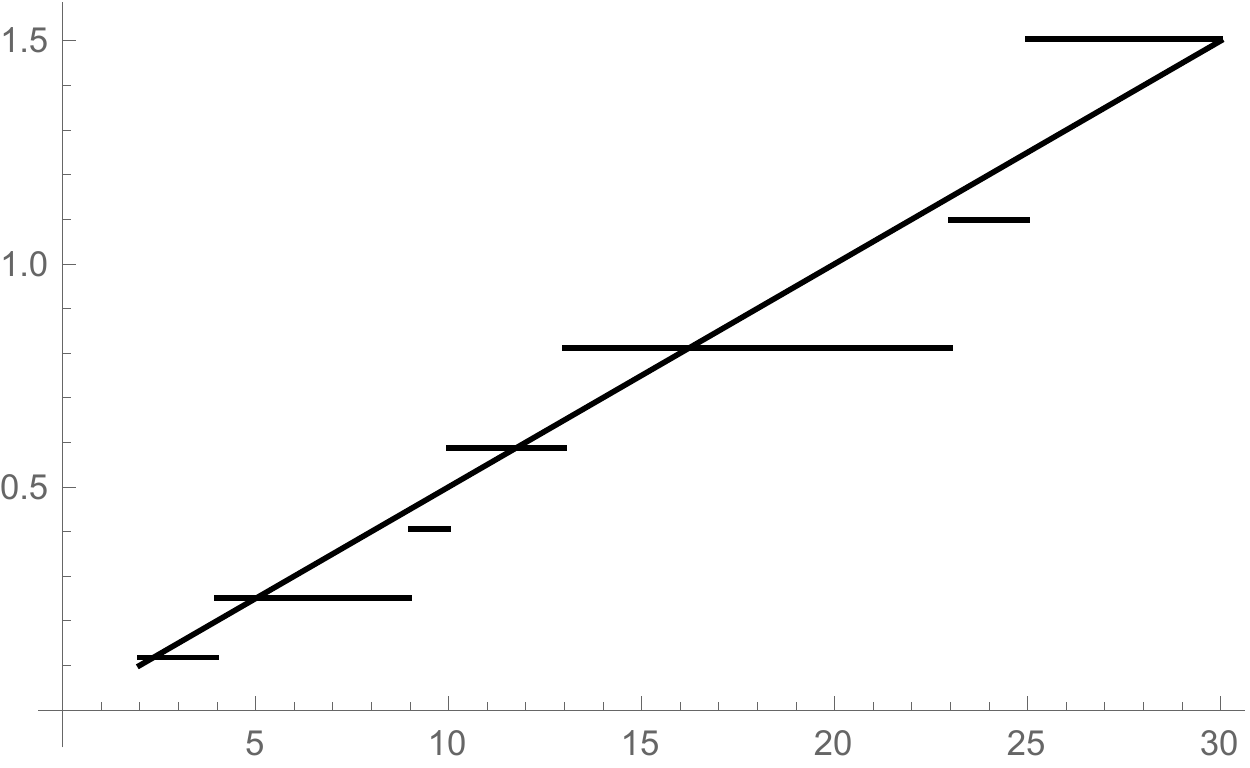}
	\caption{Plot of $-\log(1-F(x))$ for small number of publications (not greater that $30$) separately.}\label{fig5}
\end{figure}
\end{ex}

We see, that the situations for both universities are similar. The structure of observations is the same. The parts connected to not large numbers of publications are in a good agreement with geometric distribution. In both examples there are persons having abnormally large numbers of publication. Of course, the explanations of this effect given for the first example remain valued for the second as well.

\begin{ex}\label{ex3} Let $	T = \{25, 18, 50, 3, 2, 83, 60, 5, 37, 28, 14, 53, 51, 19, 47, 2, 37\}$ represent non-zero numbers of publications by members of Department of Mathematical Analysis, Charles University (data taken from Research Gate 20.05.2017). On Figures \ref{fig6} and \ref{fig7} $F(x)$ is corresponding empirical distribution function.

\begin{figure}[htp]
	\centering
	\hfil
	\includegraphics[scale=0.7]{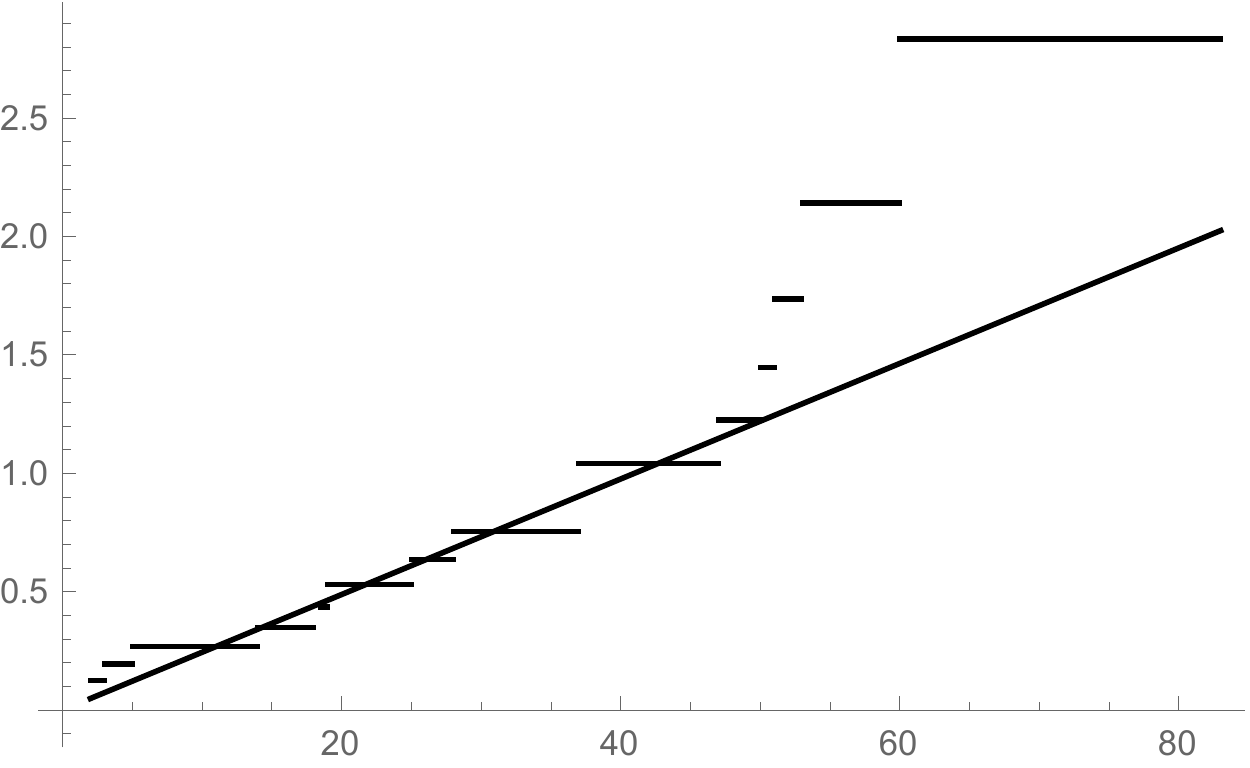}
	\caption{Plot of $-\log(1-F(x))$.}\label{fig6}
\end{figure}

\begin{figure}[htp]
	\centering
	\hfil
	\includegraphics[scale=0.7]{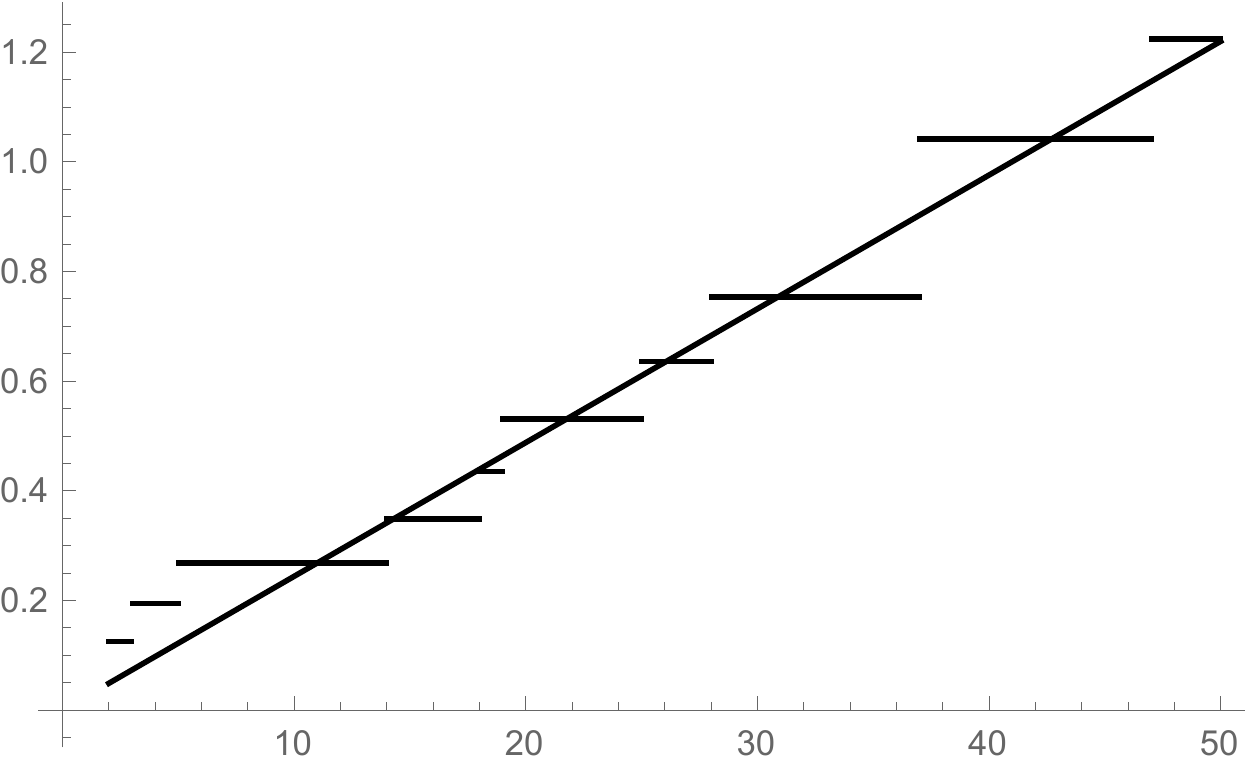}
	\caption{Plot of $-\log(1-F(x))$.}\label{fig7}
\end{figure}
\end{ex}

A part of Example \ref{ex3} seems to be very similar to that of Examples \ref{ex1} and \ref{ex2}. Namely, there is not bad agreement with geometric distribution for the case when number of publications is not too large (that is less than $50$). However, there are no persons with abnormally number of publications. For more than $50$ publications the distribution has tails lighter than in geometric case. Some of the reasons for that may be, for example, such: 
\begin{enumerate}
	\item Low qualification and experience in this field of science.
	\item Random character of the parameter $q$ from Assumption 1 (in connection, say, with different type of journals the papers were published in).
	\item Different conditions and interest to this field of science during rather long time interval passing between first and last publications.
\end{enumerate} 

\begin{ex}\label{ex4}
Let $T=\{31, 93, 7, 1, 25, 14, 9, 43, 23, 25\}$ represent non-zero numbers of publications by members of Department of Algebra, Charles University (data taken from Research Gate 20.05.2017). On Figures \ref{fig8} and \ref{fig9} $F(x)$ is corresponding empirical distribution function.
\begin{figure}[htp]
	\centering
	\hfil
	\includegraphics[scale=0.9]{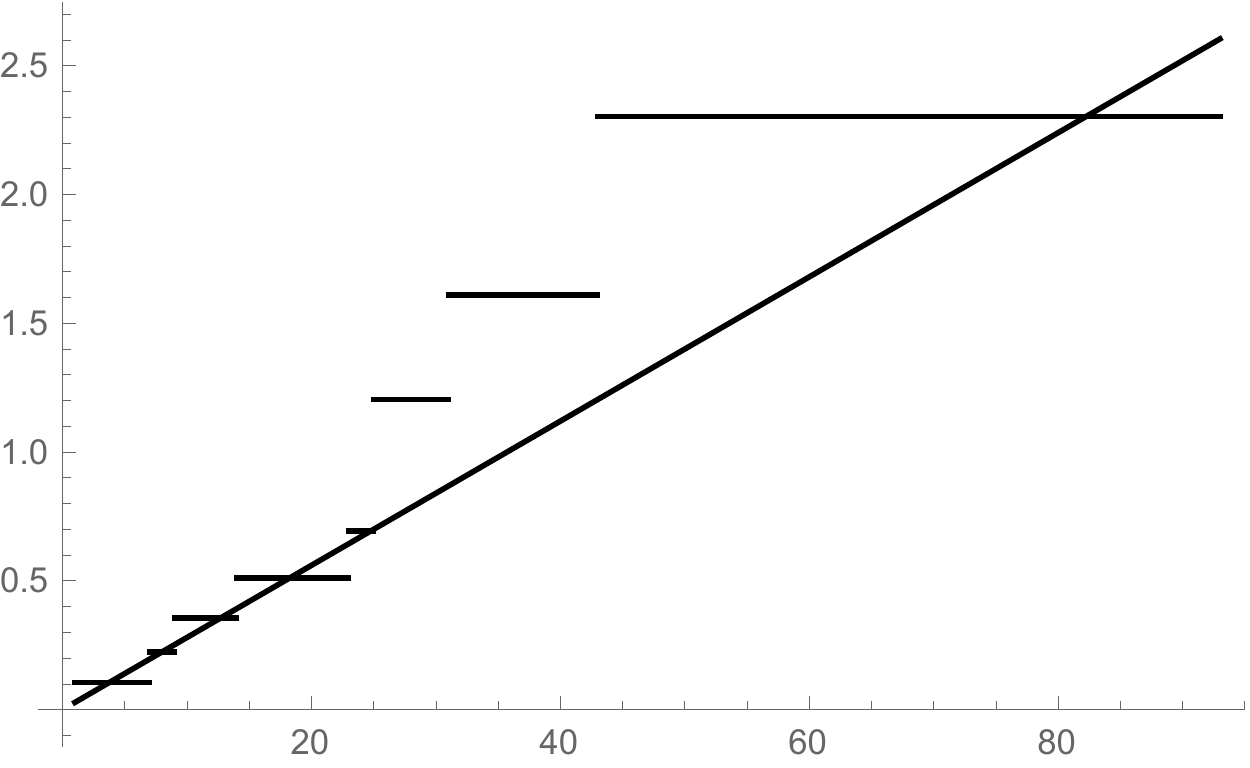}
	\caption{Plot of $-\log(1-F(x))$.}\label{fig8}
\end{figure}

\begin{figure}[htp]
	\centering
	\hfil
	\includegraphics[scale=0.9]{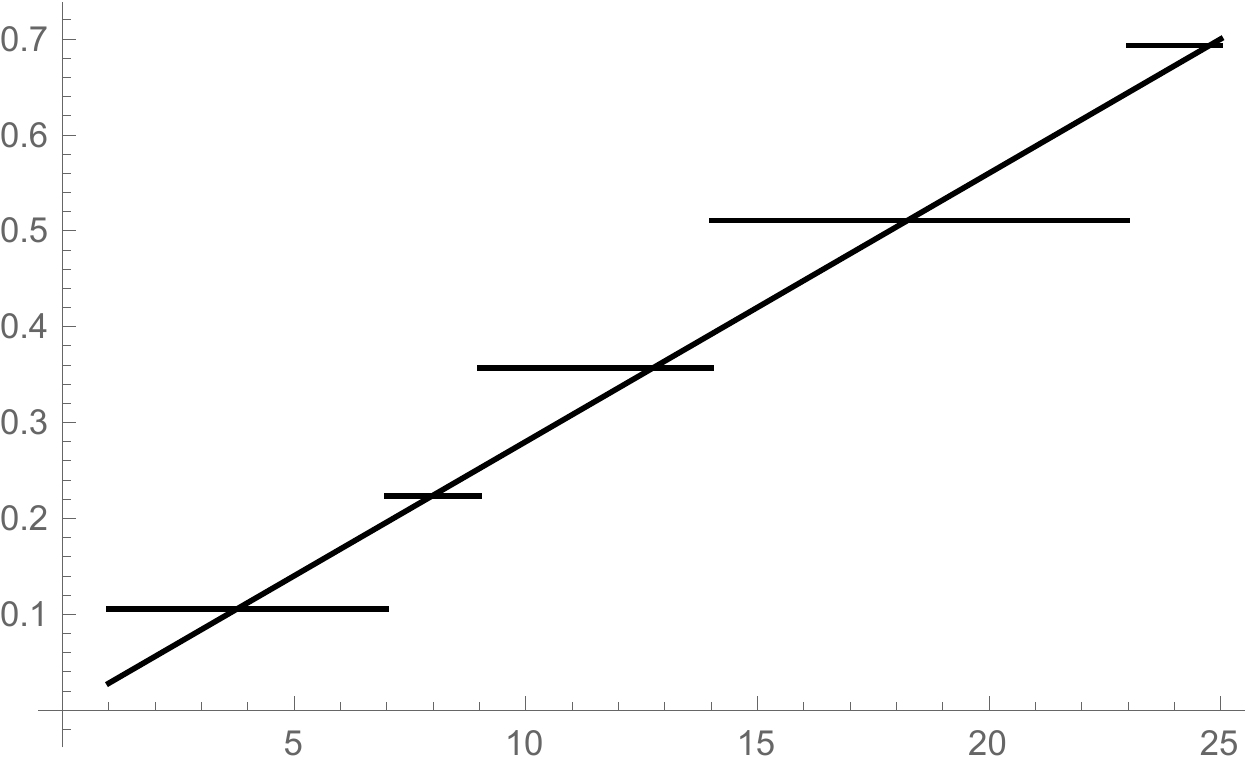}
	\caption{Plot of $-\log(1-F(x))$.}\label{fig9}
\end{figure}
\end{ex}

All the conclusions here are almost the same as in Example \ref{ex3}. We need only to change the number 50 publications by 25 of them.

\section{Some remarks of scientific ethic and pseudo-ethic}\label{sec8}
\setcounter{equation}{0}

A very interesting effect in scientific literature consists in the following. There are some papers, reported an important discovery, significance of which was not immediately realized by scientific peers. Only ten or more years after its publication did the paper get recognition, and got cited widely and
increasingly. Such papers are called ``Sleeping
Beauties” (see \cite{Raan}). For possible explanation of ``Sleeping Beauties" effect see \cite{SR5}. However, later recognition may appear just at random, in view of absence of suitable attention from the readers. Therefore, it would be not bad for the author to repeat his/her ideas in similar publication hoping for more attention to the next paper. This would be in agreement with well-known principle:
``Practice makes perfect". However, such behavior is considered by many people (especially, by editors of journals) as ``breaking scientific ethic". We think, this is a very strange consideration, because such repetition cannot produce any bad consequences for scientific work. Of course, it may produce some problems for work payment under modern principles of the fee. What is better to change: fee calculation or view on usefulness for scientific publications? We do not like to give advices, but think, here is a question for careful consideration. 

Another ``ethic" demand is author cannot send a paper in more that one journal at once. Historically, this demand really had ethical character. Thirty and more years ago publishers were obliged to do hard work to prepare any paper for publication. In addition to the authors, editors, proofreaders, compositors, and others worked on the manuscript. Therefore, a refuse to publish a paper in a journal where it had been submitted produced an essential problem for the publisher. However, now almost all preparation of the manuscript is made by the author. Therefore, it is not clear, why should authors compete for a journal and not vice versa?
We think that opposite situation will be more useful for science. At least, it will shorter time to paper publication. Only demand must be: author cannot publish this paper in more than one journal.

One more ``ethical" principle is: reviewers (and, sometimes, authors) should be anonymous for authors (respectively, for reviewers). Nowadays, the numbers of journals, authors and reviewers are very large. The probability for authors and reviewers to be in formal contact is, therefore, very low, and we do not see any reasons for anonymous character of reviewing process. Just opposite, there are some journals (one of them is ``Biology Direct") publishing not only original papers, but also corresponding reviewers reports. It gives the readers a possibility to see different opinions about the results and variety points of view on the problem. This is more essential than viewing the result itself. 

\section{Conclusions}\label{sec9}
\setcounter{equation}{0}

From proposed toy-models it is easy to see, that the role of randomness in publication and citation process may be essential. The modern methods of estimating scientific quality of publications and/or journal do not take this randomness into account. The principles of such estimation are mostly connected to business interests of publishers than to that of scientists. We are not specialists in publishing business, however, we see that modern approach to calculating rating of scientific publications has to be changed.

\section{Acknowledgment}
The work was supported by Grant GA\v{C}R 16-03708S

.

\end{document}